\documentclass[prl,aps,superscriptaddress,twocolumn,notitlepage,showpacs]{revtex4-2}
\usepackage{latexsym}
\usepackage{amsmath}  
\usepackage{amssymb}
\usepackage{graphicx}
\usepackage{caption}
\usepackage{subfigure}
\usepackage{float}
\usepackage{mathrsfs}
\usepackage{color}
\usepackage{txfonts}
\usepackage[justification=centering,
            format=plain]{caption}

\renewcommand{\raggedright}{\leftskip=0pt \rightskip=0pt plus 0cm}
\begin{document}

\title{Multidimensional Coherent Spectroscopy of Molecular Polaritons: Langevin Approach}

\author{Zhedong Zhang}
\email{zzhan26@cityu.edu.hk}
\affiliation{Department of Physics, City University of Hong Kong, Kowloon, Hong Kong SAR}
\affiliation{City University of Hong Kong, Shenzhen Research Institute, Shenzhen 518057, Guangdong, China}

\author{Xiaoyu Nie}
\affiliation{Centre for Quantum Technologies, National University of Singapore, Singapore 117543}

\author{Dangyuan Lei}
\affiliation{Department of Materials Science and Engineering, City University of Hong Kong, Kowloon, Hong Kong SAR}

\author{Shaul Mukamel}
\email{smukamel@uci.edu}
\affiliation{Department of Chemistry, University of California Irvine, Irvine, California 92697, United States}
\affiliation{Department of Physics and Astronomy, University of California Irvine, Irvine, California 92697, United States}

\date{\today}

\begin{abstract}
We present a microscopic theory for nonlinear optical spectroscopy of $N$ molecules in an optical cavity. A quantum Langevin analytical expression is derived for the time- and frequency-resolved signals accounting for arbitrary numbers of vibrational excitations. We identify clear signatures of the polariton-polaron interaction from multidimensional projections of the signal, e.g., pathways and timescales. Cooperative dynamics of cavity polaritons against intramolecular vibrations is revealed, along with a cross talk between long-range coherence and vibronic coupling that may lead to localization effects. Our results further characterize the polaritonic coherence and the population transfer that is slower. %The present analytical paradigm for time-resolved spectroscopy has promising advantages over the density matrix approach.

%Our work provides an analytical paradigm for time-resolved optical spectroscopy, promising an advantage over the density matrix approach.

\end{abstract}

\maketitle

{\it Introduction}.--Strong molecule-photon interaction has drawn considerable attention in recent study of molecular spectroscopy. New relaxation channels have been demonstrated to control fast electron dynamics and reaction activity \cite{Ebbesen_ACIE2016,Ebbesen_AC2012,Owrutsky_NatCommun2016,Huo_NatCommun2021,Spano_PRL2016,Zhou_ACSPhotonics2018,Coles_NatMater2014,Mukamel_JPCL2016,Feist_NatCommun2016}. Optical cavities create hybrid states between molecules and confined photons, known as polaritons \cite{Guebrou_PRL2012,Spano_JCP2015,Ebbesen_NatCommun2015,Rubio_PNAS2017,Haroche_RMP2001}. Theoretically, this requires a substantial generalization of quantum electrodynamics (QED) into molecules containing many more degrees of freedom than atoms and qubits.

It has been demonstrated that light in a confined geometry can significantly alter the molecular absorption and emission signals \cite{Wang_NatPhys2019,Toppari_JCP2021,Lidzey_ACSPhotonics2018}. The collective interaction between excitations of many molecules and photons is of fundamental importance, leading to interesting phenomena, e.g., superradiance and cooperative dynamics of polaritons \cite{Tavis_PR1968, Haroche_PR1982,Scully_Science2009,Spano_PRL1990,Zhang_JPCL2019,Vidal_NJP2015}. In contrast to atoms whereby superradiance and cavity polaritons are well understood, molecular polaritons are more complex in theory and experiments. This arises from the complicated couplings between electronic and nuclear degrees of freedom, which possess new challenges for optical spectroscopy. Recently the absorption and fluorescence spectra are described by Holstein-Tavis-Cummings model \cite{Spano_PRL2017,Genes_PRL2019}. Exact diagonalization of the full Hamiltonian was used to calculate the optical responses, by only taking a few vibrational excitations into account \cite{Spano_JCP2015,Spano_ARPC2014}. Here we focus on the polaritonic relaxation pathways involving the population and coherence dynamics, which are however open issues. %To bridge such a knowledge gap, the
Ultrafast spectroscopic technique has been used to monitor the dynamics of vibrational polaritons \cite{Zhang_JPCL2019,Zhou_PNAS2018,Dorfman_PNAS2018}. Time- and- frequency-gated photon-coincidence counting was employed to monitor the many-body dynamics of cavity polaritons, making use of nonlinear interferometry \cite{Dorfman_PNAS2018,Zhang_JCP2018}. Polaritons reveal the effects of strongly modifying the energy harvesting and migration in chromophore aggregates, through novel control knobs not accessible by classical light \cite{Coles_NatMater2014,Rubio_PNAS2017,Kowalewski_JPCL2016,Zhang_Optica2021,Ebbesen_ACR2016}. Elaborate nonlinear optical measurements of molecular polaritons have demonstrated unusual correlation properties \cite{Herrera_JPCL2014,Xiong_SciAdv2019,Xiong_SciAdv2021}. That calls for an extensive understanding of dark states with a high mode density \cite{Zhang_CPL2017,Zeb_ACSPhotonics2018,Feist_PRX2015,Flick_PRL2018,Lacombe_PRL2019,Kowalewski_PRA2019,Zhou_PRL2022},  %polariton-dressed nuclear dynamics \cite{Feist_PRX2015,Flick_PRL2018,Lacombe_PRL2019,Kowalewski_PRA2019}, 
nonlinearities and multiexciton correlations %in spectroscopic signals
\cite{Saurabh_JCP2016,Gu_NatCommun2021,Mukamel_JCP2016,Zhong_NatCommun2020,Mukamel_RMP1998}.

Previous spectroscopic studies of cavity polaritons were mostly based on wave function methods including nonadiabatic nuclear dynamics \cite{Kowalewski_JCP2016,Flick_JCTC2017,Kowalewski_JPCA2021}, Redfield theory and quantum chemistry simulations of low excitations \cite{Breuer_book2002,Phuc_JCP2021,Herrera_PRA2017,Vidal_JPCL2019,Tanimura_JPSJ2006,Zhou_JPCL2018}. %Vibronic coupling, however, raises a fundamental issue in cavity polaritons: absorption and emission associated with multiple phonons and photons. This leads to a strong polariton-polaron interaction that complicates the simulation of ultrafast spectroscopy. %This leads to the breakdown of perturbation paradigm, yielding a strong polariton-polaron interaction that complicates the simulation of ultrafast spectroscopy.
Absorption and emission associated with multiple phonons and optically dark states depend on a strong polariton-polaron interaction, which raises a fundamental issue in cavity polaritons and however complicates the simulation of ultrafast spectroscopy.

In this Letter, we employ a quantum Langevin theory for time-frequency-resolved coherent spectroscopy of molecular polaritons. %A general theory of time-resolved emission spectroscopy is developed for many molecules. 
Analytical solution for multidimensional third-order spectroscopic signals is developed. The results reveal multiple channels and timescales of the cooperative relaxation of polaritons, and also the trade-off with dark states. %The results unveil the polariton dynamics against the polaron-induced emitter dark states (EDSs), which may lead to a trade off with wave localization.

%The Langevin equation provides a powerful tool for describing the relaxation and radiative processes \cite{Haken_book1970,Louisell_book1990,Yan_PRA1990,Zhou_JPCL2018,Wong_JCP2021}. In such a vein, a substantial extension is essentially made in this work towards many molecules and the time-resolved regime, unveiling the polariton dynamics against the polaron-induced emitter dark states (EDSs) which may lead to a trade-off with localization of the waves. Using two sequential short pulses %with the collection direction ${\bf k}={\bf k}_{\text{probe}}$, the analytical formalism for pump-probe signal is derived, providing insights for polariton-polaron dynamics. Three sequential short pulses collecting the signal along the direction ${\bf k}=-{\bf k}_1+{\bf k}_2+{\bf k}_3$ give the multidimensional signal, revealing multiple channels and timescales for the cooperative relaxation of polaritons.

\begin{figure*}[t]
 \captionsetup{justification=raggedright,singlelinecheck=false}
\centering
\includegraphics[scale=0.39]{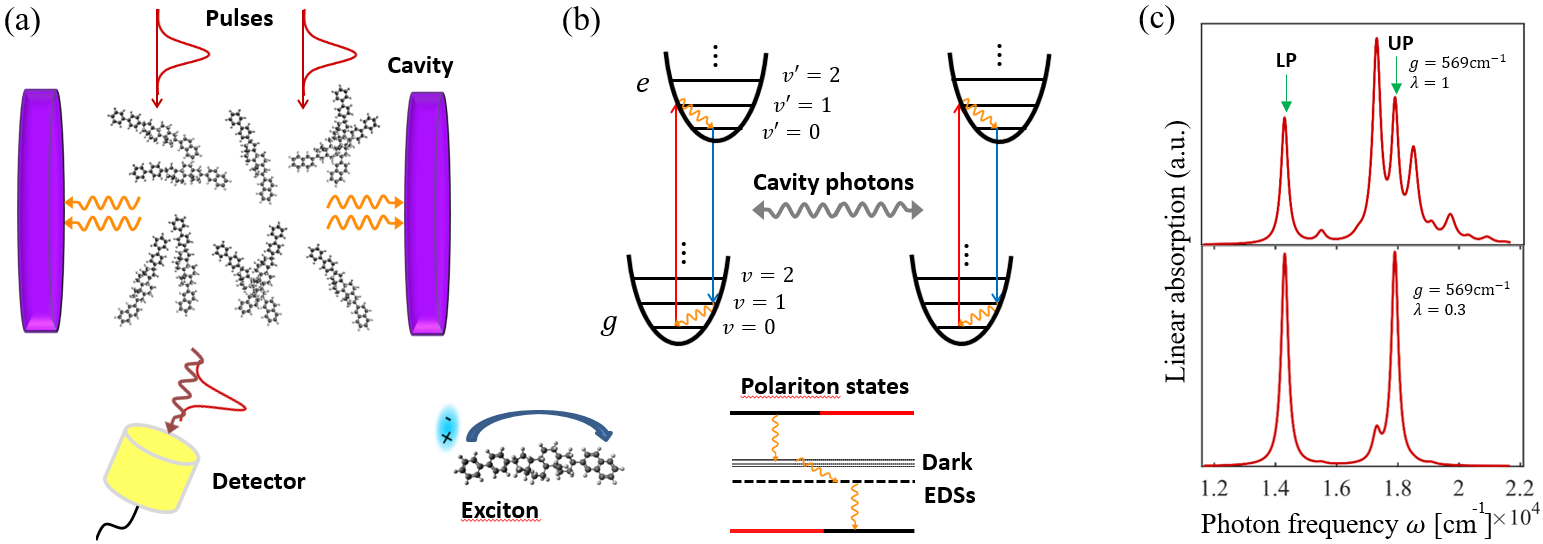}
\caption{Schematic of time-resolved spectroscopy for molecular polaritons. (a) Emission signal is collected along a certain direction, once the molecules are excited by laser pulses. (b) Exciton-photon interaction in molecules in the presence of vibronic coupling attached to individual molecule. This results in the dark states and emitter dark states (EDSs) weakly interacting with cavity, apart from the upper and lower polariton modes; Rich timescales and channels of excited-state relaxation are thus expected. (c) Linear absorption of molecular polaritons with $10$ organic molecules in an optical cavity. The parameters are taken to be $\omega_D=16113$cm$^{-1}$, $\tilde{\delta}=\delta_c=0$, $\Gamma=20$cm$^{-1}$, $\gamma=1$cm$^{-1}$, $\gamma_c=0.9$cm$^{-1}$, $\omega_{\text{v}}=1200$cm$^{-1}$, typically from cyanine dyes \cite{Peon_JPCB2013}.}
\label{F1}
\end{figure*}

{\it Langevin model for polaritons}.--Given $N$ identical molecules in an optical cavity, each has two energy surfaces corresponding to electronically ground and excited states, i.e., $|g_j\rangle$ and $|e_j\rangle\ (j=1,2,\cdots,N)$, respectively. Electronic excitations forming excitons couple to intramolecular vibrations and to cavity photons, as depicted in Fig.\ref{F1}(b), and are described by the Holstein-Tavis-Cummings Hamiltonian  %$H=\sum_{n=1}^N [H_M^{(n)} + g_n(\sigma_n^+ a + \sigma_n^- a^{\dagger})] + \delta_c a^{\dagger}a$ where
\begin{equation}
    \begin{split}
        H = \sum_{n=1}^N \left[\Delta_n\sigma_n^+ \sigma_n^- + \omega_{\text{v}} b_n^{\dagger} b_n + g_n \big(\sigma_n^+ a + \sigma_n^- a^{\dagger}\big) + \delta_c a^{\dagger}a \right]
        %H_M^{(n)} = \delta_n\sigma_n^+ \sigma_n^- + \omega_{\text{vib}}b_n^+b_n - \lambda_n\omega_{\text{vib}}\sigma_n^+ \sigma_n^- \left(b_n + b_n^{\dagger}\right)
    \end{split}
\label{HMn}
\end{equation}
where $\Delta_n=\delta - \lambda\omega_{\text{v}}(b_n + b_n^{\dagger})$ and $\delta$ denotes the detuning between excitons and external pulse field. $[\sigma_n^-,\sigma_m^+]=\sigma_n^z \delta_{nm}$. %spans Lie group $\otimes \prod_{n=1}^N \text{SU}_n(2)$.
$\sigma_n^+=|e_n\rangle\langle g_n|$ and $\sigma_n^-=|g_n\rangle\langle e_n|$ are the respective raising and lowering operators for the excitons in the $n$th molecule. $b_n$ denotes the bosonic annihilation operator of the vibrational mode with a high frequency $\omega_{\text{v}}$, in the $n$th molecule. $a$ annihilates cavity photons. %The polariton-polaron interaction emerges from the exciton-vibration-photon coupling therein. 
Each molecule has one high-frequency vibrational mode. In addition to the strong coupling to the single-longitudinal cavity mode, the molecules are subject to three temporally separated laser pulses whose electric fields $E_j(t-T_j)e^{-iv_j(t-T_j)}\ ; j=1,2,3$ described by $V(t)=\sum_{j=1}^3 \sum_{n=1}^N V_{j,n}(t) + \text{h.c.}$ with $V_{j,n}(t)=-\sigma_n^+ \Omega_j(t-T_j) e^{-i(v_j-v_3)t} e^{iv_jT_j}$ 
%\begin{equation}
%    \begin{split}
%        V(t) = - \sum_{j=1}^3 \sum_{n=1}^N \sigma_n^+ \Omega_j(t-T_j) e^{-i(v_j-v_3)t} e^{iv_jT_j} - \text{h.c.}
%    \end{split}
%\label{Vint}
%\end{equation}
%$V(t)=\sqrt{N}\sigma^+\Omega(t)+\text{h.c.}$ where $\sigma^+=N^{-1/2}\sum_{n=1}^N \sigma_n^+$ and $\Omega(t)=\sum_{j=1}^3 \Omega_j(t-T_j) e^{-i(v_j-v_3)t} e^{iv_jT_j}$. 
where $\Omega_j(t-T_j)=\mu_{eg}E_j(t-T_j)$ is the Rabi frequency with the $j$th pulse field and $\mu_{eg}$ is molecular dipole moment \cite{Condon}. The full Hamiltionian is $H(t)=H+V(t)$, which yields the quantum Langevin equations (QLEs) for $\sigma_n^-, a, b_n$.

%We have adopted the Condon approximation that $\mu_{eg}$ is independent of the nuclear coordinates \cite{Mukamel_book1999}. 

We incorporate the polaron transform via the displacement operator $D_n=e^{\lambda(b_n-b_n^{\dagger})}$ into the QLE for the dressed operator $\tilde{\sigma}_n^-=\sigma_n^-D_n^{\dagger}$. This is to involve the exciton-vibration coupling to all orders, as it is normally moderate or strong. The QLEs for operators read a matrix form %Dropping the terms $\propto$ nuclear velocity much slower than the  electrons, the QLEs for operators read a matrix form
\begin{equation}
    \begin{split}
        \dot{\text{V}} = -\hat{\text{M}}\text{V} + \text{V}^{\text{in}}(t) + i\sum_{j=1}^3 \Omega_j(t-T_j)e^{iv_j T_j}e^{-i(v_j-v_3)t}\text{W}_{\text{x}}
    \end{split}
\label{qlep4}
\end{equation}
after a lengthy algebra, where the term $\propto (b_n - b_n^{\dagger})=i\sqrt{2} p_n$ ($p_n$ is the dimensionless momentum of nuclear) has been dropped due to the nuclear velocity much lower than electrons \cite{SM}. The vector $\text{V}=[\tilde{\sigma}_1^-,\tilde{\sigma}_2^-,\cdots,\tilde{\sigma}_N^-,a]^{\text{T}}$ involves $N+1$ components, and $\text{W}_{\text{x}}=[(2n_1-1)D_1^{\dagger},\cdots,(2n_1-1)D_N^{\dagger},0]^{\text{T}}$, $n_l=\tilde{\sigma}_l^+ \tilde{\sigma}_l^-$. $\text{V}^{\text{in}}(t)=[\sqrt{2\gamma}\tilde{\sigma}_1^{-,\text{in}}(t),\cdots,\sqrt{2\gamma}\tilde{\sigma}_N^{-,\text{in}}(t),\sqrt{2\gamma_c}a^{\text{in}}(t)]^{\text{T}}$ groups the noise operators originated from exciton decay and cavity leakage. The matrix $\hat{\text{M}}$ in Eq.(\ref{qlep4}) reads
\begin{equation}
    \begin{split}
        \hat{\text{M}} = \begin{pmatrix}
                    i\tilde{\delta}+\gamma & 0 & \cdots & 0 & ig \sigma_1^z D_1^{\dagger}\\
                    0 & i\tilde{\delta}+\gamma & \cdots & 0 & ig \sigma_2^z D_2^{\dagger}\\
                    \vdots & \vdots & {} & \vdots & \vdots\\
                    0 & 0 & \cdots & i\tilde{\delta}+\gamma & ig \sigma_N^z D_N^{\dagger}\\
                    ig D_1 & ig D_2 & \cdots & ig D_N & i\delta_c + \gamma_c
                  \end{pmatrix}.
    \end{split}
\label{M}
\end{equation}
We solve for the vibration dynamics: $b_n(t)\approx e^{-(i\omega_{\text{v}}+\Gamma)t}b_n(0) + \sqrt{2\Gamma}\int_{0}^t e^{-(i\omega_{\text{v}}+\Gamma)(t-t')}b_n^{\text{in}}(t')dt'$, neglecting back influence from excitons, along the line of the stochastic Liouville equation \cite{Zhang_JPCL2019,Tanimura_JPSJ2006}. %In Eq.(\ref{qlep4}), the population $n_l$ is responsible for the emission having an opposite sign to the absorption.
Eq.(\ref{qlep4}) represents the dynamics of molecular polaritons. Perturbation theory of the molecule-field interaction $V(t)$ will be used and we will calculate two-dimensional photon emission signals by placing the detectors off the cavity axis, shown in Fig.\ref{F1}(a). These signals are governed by multipoint correlation functions of the dipole operators and the corresponding Green's functions, which are determined by the exact solution to the QLEs in Eq.(\ref{qlep4}).

{\it The polariton emission}.--We first present a general result for the emission spectrum of cavity polaritons. Subject to a probe pulse, Eq.(\ref{qlep4}) solves for the far-field dipolar radiation from molecules governed by the macroscopic polarization $P(t)=\mu_{e g}^{*} \sum_{i=1}^{N}\langle\sigma_{i ; 1}^{-}(t)\rangle$. We find the emission signal
\begin{equation}
    \begin{split}
        P_E(\omega,T) = 2i & |\mu_{eg}|^2 \sum_{i=1}^N\sum_{l=1}^N \int_{-\infty}^{\infty} \text{d}t \int_{0}^{t} \text{d}\tau\ e^{i\omega t} E(\tau-T) \\[0.15cm]
        & \ \ \times e^{-iv(\tau-T)}\langle G_{i l}(t-\tau) n_l(\tau) D_l^{\dagger}(\tau) D_i(t)\rangle
    \end{split}
\label{PAE}
\end{equation}
where $G(t)={\cal T}e^{-\int_0^t \hat{\text{M}}\text{d}t'}$ is the free propagator without pulse actions. We note that, from the dressed excited-state populations $n_l(\tau) D_l^{\dagger}(\tau)$, the cavity polaritons of molecules undergo a dynamics against the local fluctuations from polaron effect. The polaron-induced localization as a result of dark states will compete with the cooperative dynamics of polaritons. These can be visualized from the emission signal, which will thus be a real-time monitoring of polariton dynamics through pulse shaping and grating. More advanced information will be elaborated by the multidimensional projections of the signals.

{\it Linear absorption}.--Assuming $\omega_{\text{v}}/T_b\gg  1$ that applies for organic molecules at room temperature, the vibrational correlation functions can be evaluated within vacuum state. 
%the vibrations are at vacuum initially so that $\langle D_i(t)D_l^{\dagger}(\tau)\rangle = \langle 0| D_i(t)D_l^{\dagger}(\tau)|0\rangle$. 
Using Eq.(\ref{qlep4}), the absorption spectra reads $S_A(\omega)=\sum_{i,l=1}^N \sum_{m=0}^{\infty} S_m^{\lambda}\delta_{il}^m \text{Re}\left[{\cal G}_{il}(\omega-\xi_m^*)\right]$ 
%$S_A(\omega)=\text{Im}[\chi^{(1)}(\omega)]$ where the linear response function is $\chi^{(1)}(\omega) = -i \sum_{i,l=1}^N \sum_{m=0}^{\infty} S_m^{\lambda}\delta_{il}^m {\cal G}_{il}(\omega-\xi_m^*)$ 
%\begin{equation}
%    \begin{split}
%        \chi^{(1)}(\omega) = -i |\mu_{eg}|^2 \sum_{i=1}^N\sum_{l=1}^N \sum_{m=0}^{\infty} S_m^{\lambda}\delta_{il}^m {\cal G}_{il}(\omega-\xi_m^*)
%    \end{split}
%\label{SA}
%\end{equation}
and $S_m^{\lambda}=e^{-\lambda^2}\lambda^{2m}/m!$ is the Franck-Condon factor. $\xi_m=m(\omega_{\text{v}}+i\Gamma)$ and ${\cal G}_{mn}(\Omega)=\int_0^{\infty}G_{mn}(t)e^{i\Omega t}dt$ is the Fourier component of the propagator $G(t)$. $S_A(\omega)$ resolves the lower polariton (LP) and upper polariton (UP) $\omega_{\text{LP}/\text{UP}}$, EDSs $\omega_{\text{D}}+m\omega_{\text{v}}$ decoupled from cavity photons while the dark states $\omega_{\text{D}}$ are not visible. To see these closely, we assume $\gamma=\gamma_c,\ \tilde{\delta}=\delta_c=0$. The peak intensities can be thus found $S_{A}\left(\omega_{\mathrm{D}}+m \omega_{\mathrm{v}}\right)/S_{A}\left(\omega_{\mathrm{LP} / \mathrm{UP}}\right)\approx 2(\lambda^{2m}/m!)(\gamma_{\text{LP}/\text{UP}}/m\Gamma)$ and $S_{A}\left(\omega_{\mathrm{LP}/\mathrm{UP}}+m \omega_{\mathrm{v}}\right)/S_{A}\left(\omega_{\mathrm{LP} / \mathrm{UP}}\right)\approx 0$ when $N\gg 1$. The modes at $\omega_{\text{LP}/\text{UP}}+m\omega_{\text{v}}$ are hard to observe. Yet the EDSs at $\omega_D+m\omega_{\text{v}}$ may be of comparable intensities with polariton modes. Such spectral-line properties will be shown to be generally true in the time-resolved spectroscopic signals.

Fig.\ref{F1}(c,up) illustrates the absorption spectra where the LP and UP are prominent from the peaks at 14300cm$^{-1}$ and 17900cm$^{-1}$ seperated by $2g\sqrt{N}$. In between, we can observe an extra peak at $\omega_D+\omega_{\text{v}}$ supporting an EDS decoupled from cavity photons and the large oscillator strength owing to the density of states $\sim N$. Fig.\ref{F1}(c,down) shows that the EDSs are masked by the Rabi splitting for weaker vibronic coupling. This, as a benchmark to the strong-coupling case, elaborate the effect of vibronic coupling against the collective coupling to cavity photons. The localization nature of the EDSs is thus indicated from eroding the cooperativity between molecules, which will be elaborated in time-resolved spectroscopy.

{\it 2D polariton spectroscopy}.--To have multidimensional projections of the emission signal, a sequential laser pulses have to interact with the molecular polaritons. As the first two pulses create excited-state populations and coherences  $n_{l;2}(t)=\sigma_{l;1}^+(t)\sigma_{l;1}^-(t)$ where the 1st-order correction $\sigma_{l;1}^{\pm}$ is calculated from Eq.(\ref{qlep4}), we find
\begin{equation}
    \begin{split}
        n_{l;2}(t) D_l^{\dagger}(t) = & \sum_{j=1}^N \sum_{j'=1}^N\iint_0^t dt'' dt' {\cal E}_1^*(t'-T_1) {\cal E}_2(t''-T_2) \\[0.15cm]
        & \times G_{lj'}^*(t-t') G_{lj}(t-t'')D_{j'}(t')D_j^{\dagger}(t'')D_l^{\dagger}(t).
    \end{split}
\label{n2}
\end{equation}
The 3rd-order correction to the polarization follows Eq.(\ref{PAE}) when the third pulse serves as probe. Given the time-ordered pulses, the transition pathways are selective resulting from the term cancellation. Inserting Eq.(\ref{n2}) into Eq.(\ref{PAE}), we therefore proceed to the far-field polarization for the emission along the direction $\textbf{k}_{\text{I}}=-\textbf{k}_1+\textbf{k}_2+\textbf{k}_3$, i.e., $P(t)=\mu_{eg}^*\sum_{i=1}^N\langle \sigma_{i;3}^-(t)\rangle$ which yields
\begin{equation}
    \begin{split}
        P(\omega) & = 2i \sum_{i,l=1}^N\sum_{j,j'=1}^N \iiiint_{0}^{\infty} dt d\tau dt'' dt' e^{i\omega t} {\cal E}_3(\tau-T_3) \\[0.15cm]
        & \ \ \times {\cal E}_2(t''-T_2) {\cal E}_1^*(t'-T_1) \langle 0|G_{il}(t-\tau)G_{lj'}^*(\tau-t') \\[0.15cm]
        & \ \ \times G_{lj}(\tau-t'') D_{j'}(t')D_j^{\dagger}(t'')D_l^{\dagger}(\tau)D_i(t)|0\rangle
    \end{split}
\label{Pw}
\end{equation}
where the four-point correlation function of vibrations $\langle 0|D_{j'}(t')D_j^{\dagger}(t'')D_l^{\dagger}(\tau)D_i(t)|0\rangle$ has to be evaluated explicitly. The 2D signal is usually detected via a reference beam as a local oscillator interfering with the emission. This leads to the heterodyne-detected signal $S_{\text{2D}}(\Omega_3,T,\Omega_1)=\text{Im}\int_0^{\infty}E_{\text{LO}}^*(\Omega_3)P(\Omega_3) e^{i\Omega_1\tau}d\tau$ with the Fourier transform against the 1st delay $\tau=T_2-T_1$, where $E_{\text{LO}}(\Omega_3)$ is the Fourier component of the local oscillator field. In general, calculating the signal with Eq.(\ref{Pw}) is hard due to the integrals over pulse shapes. The procedures can be simplified by invoking the {\it impulsive approximation} such that the pulse is shorter than the dephasing and solvent time scales. We further consider the few-photon cavity that draws much attention in recent experiments, and notice the vibronic coupling predominately accounted by the polarons. The most significant terms may be remained, allowing the approximation $g\sigma_l^z D_l^{\dagger}\approx g D_l^{\dagger}\approx g$ in Eq.(\ref{M}). The higher-order corrections will be presented elsewhere. We obtain an analytical solution to the 2D polariton signal (2DPS), up to a real constant
\begin{equation}
    \begin{split}
        & S(\Omega_3,T,\Omega_1) = i e^{i\phi} \sum_{i,l=1}^N\sum_{j,j'=1}^N\sum_{p=1}^{N+1}\sum_{\{m\}=0}^{\infty} S_{\{m\}}^{\lambda} \delta_{j'j}^{m_1}\delta_{il}^{m_2}\delta_{jl}^{m_3}\delta_{j'l}^{m_4} \\[0.15cm]
        & \ \times \delta_{ij}^{m_5}\delta_{ij'}^{m_6}(-1)^{m_3+m_6} {\cal G}_{il}\left(\Omega_3+\xi_{m_2+m_5+m_6}\right) G_{lp}^*(T)G_{lj}(T) \\[0.15cm]
        & \ \times e^{i\xi_{m_3+m_4+m_5+m_6} T} {\cal G}_{pj'}^*\left(-\Omega_1-\xi_{m_1+m_4+m_6}\right)
    \end{split}
\label{S2D}
\end{equation}
subsequently from Eq.(\ref{Pw}), where $S_{\{m\}}^{\lambda}=\prod_{s=1}^6 S_{m_s}^{\lambda}$ and $\phi$ encodes the global phase from the four classical pulses. Details of the derivation of the signals via QLEs are given in Supplemental Material \cite{SM}. %${\cal G}_{mn}(\Omega)=\int_0^{\infty}G_{mn}(t)e^{i\Omega t}dt$ is the spectral component responsible for the line shape of the signal. 

\begin{figure*}[t]
 \captionsetup{justification=raggedright,singlelinecheck=false}
\centering
\includegraphics[scale=0.6]{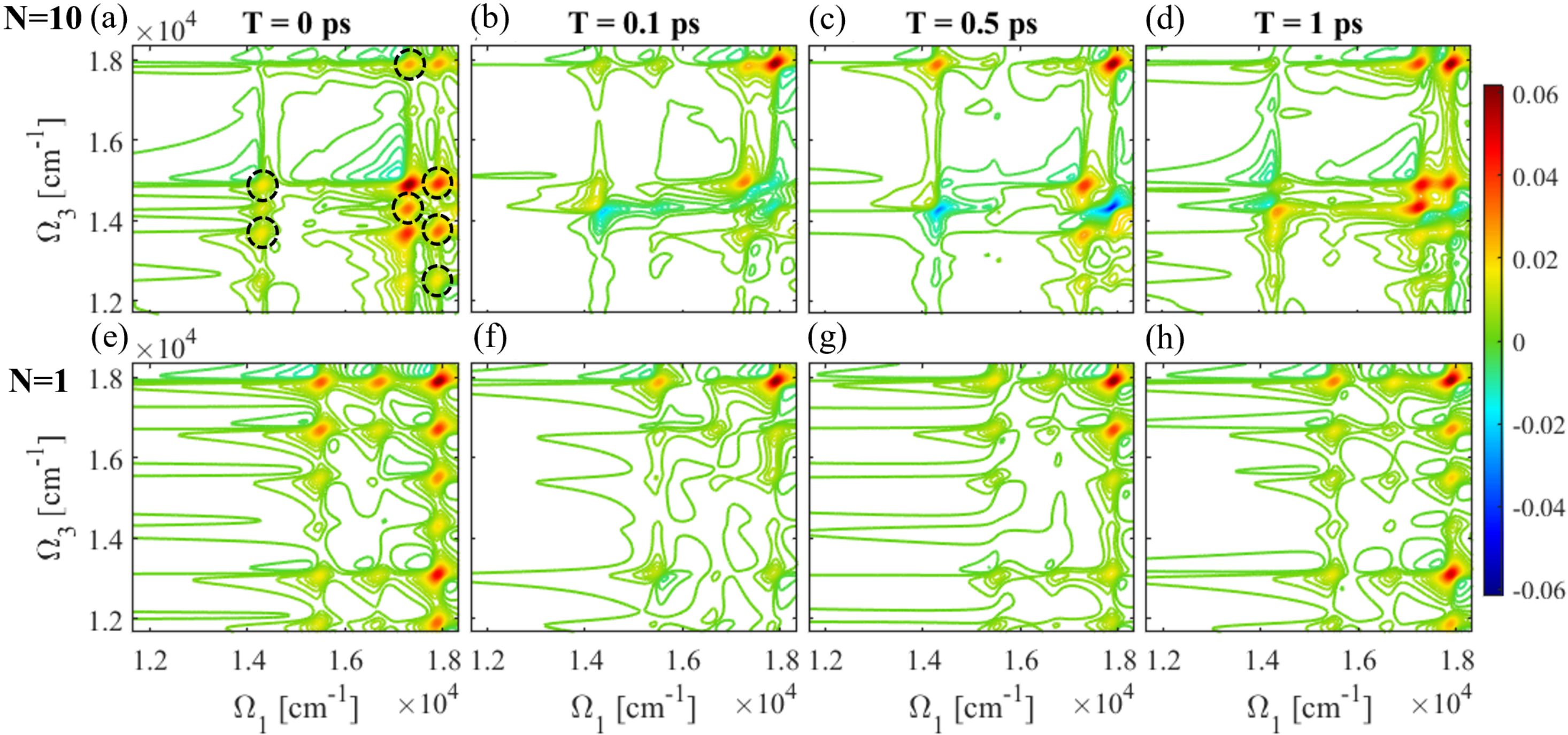}
\caption{2D time- and frequency-resolved signal for molecular polaritons where (up) $N=10$ organic molecules and (down) $N=1$ organic molecule. $T$ is varied, denoting the delay between the 2nd and the probe laser pulses. Horizontal and vertical axis are for absorption and emission frequencies, respectively. $\lambda=1$ and $g\sqrt{N}/\omega_{\text{v}}=1.5$. Other parameters are the same as Fig.\ref{F1}.}
\label{F3}
\end{figure*}

{\it Simulations}.--We have simulated the 2DPS to study polariton, exciton and polaron dynamics from the analytical solutions. We set $g\sqrt{N}/\omega_{\text{v}}=1.5$ for strong coupling. %against the polaron effect.

The lower and upper rows in Fig.\ref{F3} illustrate the 2DPS respectively for $N=1$ and $10$ molecules with fixed Rabi frequency $2g\sqrt{N}$. For 10 molecules in cavity, the signal reveals the real-time population transfer and coherence dynamics between polaritons and EDSs. The EDSs, however, cannot be resolved when one molecule coupled to cavity only. This is evident by the absence of the peaks at $\Omega_{1,3}=\omega_D\pm n \omega_{\text{v}}\  (n=0,1,2,...)$ in the lower row, compared with the upper row. The 2DPS for $N=1$ can monitor the states at $\omega_{\text{UP}}-integer\times\omega_{\text{v}}$ and their population transfer as well as coherence with the polariton states, as seen from the variation of the cross peaks, for example, at $(\Omega_1=\omega_{\text{UP}}-n\omega_{\text{v}},\Omega_3=\omega_{\text{UP}}-m\omega_{\text{v}}), m\neq n$ with the time delay $T$. We will present more details about the polariton dynamics against the EDSs next.

Fig.\ref{F3}(a) shows the 2D signal at $T=0$, from which the LP and UP states can be observed, evident by the two diagonal peaks at $\omega\pm g\sqrt{N}$. The cross peaks may result from the coherence and the polariton-polaron coupling, since energy and dephasing are not allowed at $T=0$. The former is due to the broadband nature of pump pulses while the latter is responsible for the change of phonon numbers associated with optical transitions. To have a closer look, we notice the states at $\Omega_1=14300, 17300$ and $17900$cm$^{-1}$, after the absorbing energy from the pump pulse. These agree with the absorbance in Fig.\ref{F1}(c,up). The cross peaks imposing $\Omega_1-\Omega_3=integer\times \omega_{\text{v}}$ indicates the population of the EDSs which decouple from cavity photons and radiate phonons when emitting photons, evident by those at $(\Omega_1=17300\text{cm}^{-1}, \Omega_3=14900\text{cm}^{-1})$ and $(\Omega_1=17300\text{cm}^{-1}, \Omega_3=13700\text{cm}^{-1})$. The EDSs erode eroding molecular cooperativity and are highly degenerated, having the frequency $\omega_D \pm n \omega_{\text{v}}; n=1,2,...$. The cross peaks with $\Omega_1-\Omega_3\neq integer\times \omega_{\text{v}}$ as circled in Fig.\ref{F3}(a), come from the coherences quantified by the off-diagonal elements of the density matrix. For instance, the one at $(\Omega_1=17900\text{cm}^{-1}, \Omega_3=14900\text{cm}^{-1})$ resolves the coherence $\rho_{\omega_D-\omega_{\text{v}},\omega_{\text{UP}}}(T)$. The peaks associated with EDSs in both absorption spectrum and the 2D signal $S_{\text{2D}}(\Omega_3,0,\Omega_1)$ manifest in the steady state of molecular ensemble the role of local fluctuations that erodes the cooperative motion of exciton polaritons. This incredibly differs from previous studies showing the dynamical breakdown of the polariton-induced cooperativity of molecules by solvent motion which does not affect the polaritons significantly in the absorption spectrum, apart from the line broadening \cite{Zhang_JPCL2019}.

When the delay between the 2nd and probe pulses varies, Fig.\ref{F3}(b) show the fast decay of the coherence produced from the first two broadband pulses. This can be seen prominently from the decreasing intensities of the cross peaks with $\Omega_1-\Omega_3\neq integer\times \omega_{\text{v}}$, compared to Fig.\ref{F3}(a). From Fig.\ref{F3}(b), nevertheless, one can observe the cross peak at $(\Omega_1=17900\text{cm}^{-1}, \Omega_3=14900\text{cm}^{-1})$ whose intensity increases after a rapid decay with the delay $T$. This describes the down-hill energy transfer from UP to the EDS $\omega=\omega_D - \omega_{\text{v}}=14900$cm$^{-1}$, following a fast dephasing. An energy transfer to LP is elucidated as well, from the growth of the cross peak at $(\Omega_1=17900\text{cm}^{-1}, \Omega_3=14300\text{cm}^{-1})$. Similarly, the energy transfer pathway from the EDS $\omega=\omega_D+\omega_{\text{v}}=17300\text{cm}^{-1}$ to the LP can be observed within about 300fs. Observing the weak intensity along with a slow growth at the cross peaks $(\Omega_1=17900\text{cm}^{-1}, \Omega_3=\omega_D-n\omega_{\text{v}})$ in Fig.\ref{F3}(b,c), the molecular system is slightly localized from the UP state within $\sim 500$fs because of the weak populations of EDSs. Within a longer timescale $T>500$fs, Fig.\ref{F3}(c,d) evidence that the energy flowing from UP and the state $\omega=\omega_D+\omega_{\text{v}}=17300\text{cm}^{-1}$ to the EDS $\omega=\omega_D - \omega_{\text{v}}=14900\text{cm}^{-1}$ dominates, leading to strong localization of the UP state. This can be neatly understood from the Fermi-Golden rule $\Gamma_{i\rightarrow f} = 2\pi |\langle f|V|i\rangle|^2 N_i D_f$ by noting a larger number of EDSs than polaritons. From the slice $\Omega_1=\omega_D+\omega_{\text{v}}=17300\text{cm}^{-1}$ in which the system can be pumped to the state at $\omega_D+\omega_{\text{v}}$ by the first two pulses, the system tends to be alternatively delocalized within a longer timescale $T>500$fs, seen from Fig.\ref{F3}(c,d) showing the strong peak intensity at LP state that indicates the population transferred considerably.

Moreover, one notes in Fig.\ref{F3} that most of the cross peaks appear below the diagonal. This results from the low temperature assumed in our model, i.e., $ \omega_{\text{v}}/T_b\gg 1$ so that the vibration modes are at vacuum initially before the pulse actions. During the first 250fs, the fast decay of the cross peak at $(\Omega_1=14300\text{cm}^{-1}, \Omega_3=14900\text{cm}^{-1})$ monitors the dephasing of the coherence $\rho_{\omega_D-\omega_{\text{v}},\omega_{\text{LP}}}(T)$. The population transfer from the LP state to the EDS $\omega=\omega_D-\omega_{\text{v}}=14900\text{cm}^{-1}$ follows when the delay $T$ becomes longer, as indicated from the cross peak at $(\Omega_1=14300\text{cm}^{-1}, \Omega_3=14900\text{cm}^{-1})$ that increases within about 750fs. Besides, the cross peak at $(\Omega_1=14300\text{cm}^{-1}, \Omega_3=17900\text{cm}^{-1})$ shows up weakly within about 500fs, as depicted in Fig.\ref{F3}(c). A small portion of energy transferred from the LP state to the UP state is thus indicated. This may be attributed to the rapid energy exchange between the polariton and vibration modes, yielding a cascading migration of populations between two polariton modes within a short timescale. In longer timescales, such behavior is expected to deplete.

{\it Relation to the pump-probe signal}.--The pump-probe signal can be readily obtained by letting $T_1=T_2$ in Eq.(\ref{Pw}) and accounting for the non-rephasing component. The signal reads $S_{\text{pp}}(\omega,T)=\text{Im}[E_3^*(\omega)P(\omega)]$ and some algebra gives
\begin{equation}
    \begin{split}
        S_{\text{pp}}(\omega,T) = & \sum_{i,l=1}^N\sum_{j,j'=1}^N \sum_{\{m\}=0}^{\infty} S_{\{m\}}^{\lambda}  \delta_{il}^{m_1} \left(\delta_{j'l} - \delta_{jl}\right)^{m_2} \left(\delta_{ij} - \delta_{ij'}\right)^{m_3}\\[0.15cm]
        & \times \text{Re}\left[{\cal G}_{il}\left(\omega+\xi_{m_1+m_3}\right)G_{lj'}^*(T)G_{lj}(T) e^{i\xi_{m_2+m_3}T}\right]
        %P(\omega) = & 2i \sum_{i,l=1}^N\sum_{j,j'=1}^N \iiiint_{0}^{\infty} dt d\tau dt'' dt'\ e^{i\omega t} {\cal E}_3(\tau-T_3) \\[0.15cm]
        %& \times {\cal E}_1(t''-T_1) {\cal E}_1^*(t'-T_1) \langle 0|G_{il}(t-\tau)G_{lj'}^*(\tau-t') \\[0.15cm]
        %& \times G_{lj}(\tau-t'') D_{j'}(t')D_j^{\dagger}(t'')D_l^{\dagger}(\tau)D_i(t)|0\rangle
    \end{split}
\label{Pwpp}
\end{equation}
under the {\it impulsive approximation}. The broadband nature of the ultrashort pulses smears out the mode selectivity in the absorption of molecular polaritons, whereas the time grating makes the emission spectrally resolved. Similar as the 2DPS, the LP and UP modes separated by $2g\sqrt{N}$ as well as the EDSs at $\omega=\omega_D-m\omega_{\text{v}}$ can be resolved in $S_{\text{pp}}(\omega,T)$. As varying the time delay, the spectral-line intensity $S_{\text{pp}}(\omega_{\text{UP}/\text{LP}},T)$ shows a phase difference from the $S_{\text{pp}}(\omega_{\text{D}}-m\omega_{\text{v}},T)$, associated with different damping rates that are responsible for the incoherent channels of relaxations.  %One would expect in further a $\pi$ phase difference between the time-evolving dynamics of UP/LP and EDSs, for properly-chosen parameters. The polariton featuring the extended waves are thus trading off with the localization nature of the EDSs. %as expected from the spectral-line intensities varying with the delay $T$. 
The low spectral resolution with the absorption process, however, makes the pump-probe signal not capable of unveiling advanced information about the polariton dynamics and dark states, e.g., relaxation pathways and timescales.  %The 2D signal is essential in this regard.
More details are given in SM \cite{SM}.

{\it Summary and outlook}.--The microscopic theory of multidimensional spectroscopy for the molecular polaritons was developed, using the quantum Langevin equation capable of polariton-polaron interactions with high excitation number. Rich information about the fast-evolving dynamics of polaritons and dark states and their couplings can be readily visualized in the 2DPS, i.e., pathways and timescales. Our work manifests the ultrafast polariton-polaron interaction in molecules, resolving the EDSs against the polariton dynamics. This falls into a different category from the cavity QED for atoms, in which no relaxation between superradiant and subradiant states could be observed \cite{Pavolini_PRL1985,Kaiser_PRL2016,Lukin_Science2018}. Understanding the collective nature of molecular polaritons is significant for the community to gain more details about the diverse phenomena afforded by the polaritons in complex materials. This knowledge may help in the design of cavity-coupled heterostructures in visible regime and in polariton chemistry.

\vspace{0.15cm}
Z.D.Z. gratefully acknowledges the support of ARPC-CityU new research initiative/infrastructure support from central (No. 9610505), the Early Career Scheme from Hong Kong Research Grants Council (No. 21302721) and the National Science Foundation of China (No. 12104380). D.L. gratefully acknowledges the support of the National Science Foundation of China, the Excellent Young Scientist Fund (No. 62022001). S.M. thanks the support of the National Science Foundation (No. CHE-1953045) and of the U.S. Department of Energy, Office of Science, Basic Energy Sciences, under Award No. DE-SC0022134.

\end{document}

\title{Multidimensional Coherent Spectroscopy of Molecular Polaritons: Langevin Approach}

\author{Zhedong Zhang}
\email{zzhan26@cityu.edu.hk}
\affiliation{Department of Physics, City University of Hong Kong, Kowloon, Hong Kong SAR}
\affiliation{City University of Hong Kong, Shenzhen Research Institute, Shenzhen 518057, Guangdong, China}

\author{Xiaoyu Nie}
\affiliation{Centre for Quantum Technologies, National University of Singapore, Singapore 117543}

\author{Dangyuan Lei}
\affiliation{Department of Materials Science and Engineering, City University of Hong Kong, Kowloon, Hong Kong SAR}

\author{Shaul Mukamel}
\email{smukamel@uci.edu}
\affiliation{Department of Chemistry, University of California Irvine, Irvine, California 92697, United States}
\affiliation{Department of Physics and Astronomy, University of California Irvine, Irvine, California 92697, United States}

\date{\today}

\begin{abstract}
The math details in addition to the conclusions delivered in main text are provided. Meanwhile, some supplemental results on multidimensional spectroscopy of cavity polaritons are shown, in support of the main text.
\end{abstract}

\maketitle

\section{LANGEVIN MODEL FOR POLARON POLARITONS}
Given $N$ identical molecules in an optical cavity, the Holstein-Tavis-Cummings Hamiltonian reads
\begin{equation}
    \begin{split}
        H_0 = \sum_{n=1}^N \left[\delta_n\sigma_n^+ \sigma_n^- + \omega_{\text{vib}}b_n^{\dagger} b_n - \lambda_n\omega_{\text{vib}}\sigma_n^+ \sigma_n^- \left(b_n + b_n^{\dagger}\right) + g_n\left(\sigma_n^+a + \sigma_n^-a^{\dagger}\right) \right] + \delta_c a^{\dagger}a
    \end{split}
\label{Hfull}
\end{equation}
in the rotating frame of photons, where $\sigma_n^+$ is the raising operator leading to electronic excitation of the $n$th molecule. $b_n$ denotes the annihilation operator of the vibrational mode having the frequency $\omega_{\text{vib}}$. $a$ annihilates a photon in cavity and $\delta_n=\omega_n-v$ denotes the detuning between excitons and external classical field. In the presence of the classical laser pulses, the full Hamiltonian is $H=H_0 + V(t)$ and
\begin{equation}
    \begin{split}
        V(t) = -\sum_{n=1}^N \sum_{j=1}^3 \left[\sigma_n^+ \Omega_j(t-T_j) e^{-i(v_j-v_3)t} e^{iv_jT_j} + \sigma_n^- \Omega_j^*(t-T_j) e^{i(v_j-v_3)t} e^{-iv_jT_j}\right]
    \end{split}
\label{V}
\end{equation}
with the Rabi frequency $\Omega_j(t-T_j)=\mu_{eg}E_j(t-T_j)$. The quantum Langevin equations (QLEs) for operators are of the form
\begin{subequations}
    \begin{align}
        & \dot{\sigma}_n^- = -\left(i\delta + \gamma\right)\sigma_n^- + i\lambda\omega_{\text{vib}}\sigma_n^- \left(b_n + b_n^{\dagger}\right) - ig\sigma_n^z a -i\sum_{j=1}^3\sigma_n^z\Omega_j(t-T_j) e^{-i(v_j-v_3)t} e^{iv_jT_j} + \sqrt{2\gamma}\sigma_n^{-,\text{in}}(t) \label{sigmaj} \\[0.15cm]
        & \dot{a} = -(i\delta_c + \gamma_c)a - ig\sum_{n=1}^N\sigma_n^- + \sqrt{2\gamma_c}a^{-,\text{in}}(t) \label{a} \\[0.15cm]
        & \dot{b}_n = -(i\omega_{\text{vib}}+\Gamma)b_n + i\lambda\omega_{\text{vib}}\sigma_n^+\sigma_n^- + \sqrt{2\Gamma}b_n^{-,\text{in}}(t) \label{bj}
    \end{align}
\end{subequations}
where the noise operators $\sigma_n^{-,\text{in}}(t),a^{-,\text{in}}(t),b_n^{-,\text{in}}(t)$ obey the fluctuation-dissipation relation
\begin{subequations}
    \begin{align}
        & \langle b_j^{\text{in},\dagger}(t)b_{j'}^{\text{in}}(t')\rangle = \bar{n}_{\text{th}}\delta_{jj'}\delta(t-t'),\quad \langle b_j^{\text{in}}(t)b_{j'}^{\text{in},\dagger}(t')\rangle = (\bar{n}_{\text{th}}+1)\delta_{jj'}\delta(t-t')\\[0.15cm]
        & \langle a^{\text{in},\dagger}(t) a^{\text{in}}(t')\rangle = 0,\quad \langle a^{\text{in}}(t) a^{\text{in},\dagger}(t')\rangle = \delta(t-t') \\[0.15cm]
        & \langle \sigma_j^{\text{in},+}(t) \sigma_{j'}^{\text{in},-}(t')\rangle = 0,\quad \langle \sigma_j^{\text{in},-}(t) \sigma_{j'}^{\text{in},+}(t')\rangle = \langle\sigma_j^- \sigma_j^+ \rangle \delta_{jj'}\delta(t-t').
    \end{align}
\end{subequations}

We apply the polaron transform to diagonalize the moelcular Hamiltonian apart from the photon degrees of freedom
\begin{equation}
    \begin{split}
        D = \otimes\prod_{n=1}^N D_n,\quad D_n = e^{\lambda(b_n-b_n^{\dagger})}.
    \end{split}
\label{polaron}
\end{equation}
Defining the dressed lowering operator: $\tilde{\sigma}_n^-=\sigma_n^-D_n^{\dagger}$, one has
\begin{equation}
    \dot{D}_n = \lambda D_n \left[i\omega_{\text{vib}}\left(b_n+b_n^{\dagger} - 2\lambda \sigma_n^+\sigma_n^-\right) - \Gamma\left(b_n^{\dagger}-b_n\right) + \sqrt{2\Gamma}\left(b_n^{\text{in},\dagger}(t)-b_n^{\text{in}}(t)\right) \right].
\label{Dn}
\end{equation}
Inserting Eq.(\ref{Dn}) into Eq.(\ref{sigmaj}) and taking the time derivative of $\tilde{\sigma}_n^-$, some algebra gives
\begin{equation}
  \begin{split}
    %\dot{\tilde{\sigma}}_n^- = - \left[i(\delta-2\lambda^2 \omega_{\text{vib}})+\gamma\right] \tilde{\sigma}_n^-  - ig \sigma_n^z a D_n^{\dagger} - i\sum_{j=1}^3 \sigma_n^z D_n^{\dagger} \Omega_j(t-T_j)e^{-i(v_j-v_3)t} e^{iv_j T_j} + \sqrt{2\gamma} \tilde{\sigma}_n^{-,\text{in}}(t)
    \dot{\tilde{\sigma}}_n^- = - & \left[i(\delta-2\lambda^2 \omega_{\text{vib}})+\gamma\right] \tilde{\sigma}_n^- - ig \sigma_n^z a D_n^{\dagger} - i\sum_{j=1}^3 \sigma_n^z D_n^{\dagger} \Omega_j(t-T_j)e^{-i(v_j-v_3)t} e^{iv_j T_j} \\[0.15cm]
    & \qquad\qquad\quad + \lambda\Gamma \left(b_n^{\dagger}-b_n\right)\tilde{\sigma}_n^- - \lambda\sqrt{2\Gamma}\left[b_n^{\text{in},\dagger}(t)-b_n^{\text{in}}(t)\right]\tilde{\sigma}_n^- + \sqrt{2\gamma} \tilde{\sigma}_n^{-,\text{in}}(t)
  \end{split}
\label{sigmat}
\end{equation}
and $\tilde{\sigma}_n^{-,\text{in}}(t)=\sigma_n^{-,\text{in}}D_n^{\dagger}$. It turns out that the time dynamics of the population $\tilde{\sigma}_n^+ \tilde{\sigma}_n^-$ eliminates the contribution from the term $\lambda\sqrt{2\Gamma}\left[b_n^{\text{in},\dagger}(t)-b_n^{\text{in}}(t)\right]$ such that
\begin{equation}
    \dot{n}_l \rightarrow - \lambda\sqrt{2\Gamma}\left[b_l^{\text{in},\dagger}(t)-b_l^{\text{in}}(t) + \text{h.c.}\right] n_l = 0.
\end{equation}
The term $\lambda\sqrt{2\Gamma}\left[b_n^{\text{in},\dagger}(t)-b_n^{\text{in}}(t)\right]$ can be properly dropped, for a good approximation. One observes that the nuclear velocity $\propto i( b_n^{\dagger}-b_n)$ which is much slower than electrons and holes. We may further drop the term correspondingly and neglect the back influence from exciton to nuclear motion, where the latter recasts Eq.(\ref{bj}) into $\dot{b}_n \approx -(i\omega_{\text{vib}}+\Gamma)b_n + \sqrt{2\Gamma}b_n^{-,\text{in}}(t)$. We then solve for the vibration dynamics
\begin{equation}
    b_n(t) \approx e^{-(i\omega_{\text{vib}}+\Gamma)t}b_n(0) +  \sqrt{2\Gamma} \int_0^t e^{-(i\omega_{\text{vib}}+\Gamma)(t-t')}b_n^{-,\text{in}}(t')dt'
\label{bns}
\end{equation}
which possess
\begin{equation}
    [b_n(t),b_m^{\dagger}(t')] = \begin{cases}
                                   e^{-(i\omega_{\text{vib}}+\Gamma)(t-t')} \delta_{nm}, & \text{for } t\ge t'\\[0.15cm]
                                   e^{(i\omega_{\text{vib}}-\Gamma)(t'-t)} \delta_{nm}, & \text{for } t < t'.
                                 \end{cases}
\label{bnc}
\end{equation}
We recast the QLEs into
\begin{subequations}
    \begin{align}
      & \dot{\tilde{\sigma}}_l^- = -(i\tilde{\delta}+\gamma)\tilde{\sigma}_l^- - ig \sigma_l^z a D_l^{\dagger} +  i\sum_{j=1}^3(2n_l-1)D_l^{\dagger}\Omega_j(t-T_j) e^{-i(v_j-v_3)t} e^{iv_jT_j} + \sqrt{2\gamma}\tilde{\sigma}_l^{-,\text{in}}(t)\\[0.15cm]
      & \dot{a} = -(i\delta_c + \gamma_c)a - ig\sum_{n=1}^N\tilde{\sigma}_n^- D_n + \sqrt{2\gamma_c}a^{-,\text{in}}(t),\quad \tilde{\delta} \equiv \delta - 2\lambda^2 \omega_{\text{vib}}
    \end{align}
\label{qlep2}
\end{subequations}
which can be reformed into a compact form
\begin{equation}
    \begin{split}
        \dot{\text{V}} = -\text{M}\text{V} + \text{V}^{\text{in}}(t) + i\sum_{j=1}^3 \Omega_j(t-T_j)e^{iv_j T_j}e^{-i(v_j-v_3)t}\text{W}_{\text{x}}
    \end{split}
\label{qlep4}
\end{equation}
where the related matrices are
\begin{equation}
    \begin{split}
      \text{V} = \begin{pmatrix}
                   \tilde{\sigma}_1^-\\
                   \tilde{\sigma}_2^-\\
                   \vdots\\
                   \tilde{\sigma}_N^-\\
                   a
                 \end{pmatrix},\
       \hat{\text{M}} = \begin{pmatrix}
                    i\tilde{\delta}+\gamma & 0 & \cdots & 0 & ig\sigma_1^z D_1^{\dagger} \\
                    0 & i\tilde{\delta}+\gamma & \cdots & 0 & ig\sigma_2^z D_2^{\dagger} \\
                    \vdots & \vdots & {} & \vdots & \vdots\\
                    0 & 0 & \cdots & i\tilde{\delta}+\gamma & ig\sigma_N^z D_N^{\dagger} \\
                    ig D_1 & ig D_2 & \cdots & ig D_N & i\delta_c + \gamma_c
                  \end{pmatrix},\
        \text{V}^{\text{in}}(t) = \begin{pmatrix}
                                    \sqrt{2\gamma}\tilde{\sigma}_1^{-,\text{in}}(t)\\
                                    \sqrt{2\gamma}\tilde{\sigma}_2^{-,\text{in}}(t)\\
                                    \vdots\\
                                    \sqrt{2\gamma}\tilde{\sigma}_N^{-,\text{in}}(t)\\
                                    \sqrt{2\gamma_c}a^{\text{in}}(t)
                                  \end{pmatrix},\
          \text{W}_\text{x} = \begin{pmatrix}
                                (2n_1-1)D_1^{\dagger}\\
                                (2n_2-1)D_2^{\dagger}\\
                                \vdots\\
                                (2n_N-1)D_N^{\dagger}\\
                                0
                              \end{pmatrix}.
    \end{split}
\label{matrix}
\end{equation}

We find following identities for the vibrations \cite{Mukamel_book1999,Scully_book1997}
\begin{equation}
  \begin{split}
    & e^{\lambda[b_n^{\dagger}(t)-b_n(t)]} = e^{-\lambda^2/2}e^{\lambda b_n^{\dagger}(t)} e^{-\lambda b_n(t)} = e^{\lambda^2/2}e^{-\lambda b_n(t)} e^{\lambda b_n^{\dagger}(t)}\\[0.15cm]
    & e^{-\lambda b_n(t)}|0\rangle = |0\rangle
  \end{split}
\end{equation}
and $e^A e^B = e^B e^A e^{[A,B]}$, given $[[A,B],A]=[[A,B],B]=0$ from the Baker-Hausdorff formula. Eq.(\ref{bns}) thus enables us to calculate the vibration correlation functions. For instance, the two-point correlator reads
\begin{equation}
  \begin{split}
    e^{\lambda^2}\langle 0|D_i^{\dagger}(t)D_j(t')|0\rangle & = \langle 0|e^{\lambda b_i(t)} e^{\lambda b_j^{\dagger}(t')}|0\rangle\\[0.15cm]
    & = \langle 0|e^{\lambda b_j^{\dagger}(t')}e^{\lambda b_i(t)} e^{\lambda^2[b_i(t),b_j^{\dagger}(t')]}|0\rangle\\[0.15cm]
    & = \theta(t-t')\text{exp}\left[\delta_{ij}\lambda^2 e^{-(i\omega_{\text{vib}}+\Gamma)(t-t')}\right] + \theta(t'-t)\text{exp}\left[\delta_{ij}\lambda^2 e^{(i\omega_{\text{vib}}-\Gamma)(t'-t)}\right].
  \end{split}
\end{equation}
%The calculation of four-point correlators follows the same strategy with more redundancy of algebra, which will be entailed later on.
The four-point correlator is calculated in further
\begin{equation}
    \begin{split}
        & e^{2\lambda^2}\langle 0|D_{j'}(t)D_j^{\dagger}(t')D_l^{\dagger}(t'')D_i(t''')|0\rangle\\[0.15cm]
        & = \langle 0|e^{-\lambda b_{j'}(t)}e^{-\lambda b_j^{\dagger}(t')}e^{\lambda b_j(t')}e^{-\lambda b_l^{\dagger}(t'')}e^{\lambda b_l(t'')}e^{\lambda b_i^{\dagger}(t''')}|0\rangle\\[0.15cm]
        & = e^{\lambda^2 [b_{j'}(t),b_j^{\dagger}(t')]} e^{\lambda^2 [b_l(t''),b_i^{\dagger}(t''')]} \langle 0| e^{-\lambda b_{j'}(t)} e^{\lambda b_j(t')} e^{-\lambda b_l^{\dagger}(t'')} e^{\lambda b_i^{\dagger}(t''')}|0\rangle\\[0.15cm]
        & = e^{\lambda^2 [b_{j'}(t),b_j^{\dagger}(t')]} e^{\lambda^2 [b_l(t''),b_i^{\dagger}(t''')]} e^{-\lambda^2 [b_j(t'),b_l^{\dagger}(t'')]} e^{\lambda^2 [b_{j'}(t),b_l^{\dagger}(t'')]} e^{\lambda^2 [b_j(t'),b_i^{\dagger}(t''')]}
        e^{-\lambda^2 [b_{j'}(t),b_i^{\dagger}(t''')]}
    \end{split}
\label{vibcorr}
\end{equation}
which is general. For the segment $t\le t'\le t''\le t'''$, it reads
\begin{equation}
    \begin{split}
        & e^{2\lambda^2}\langle 0|D_{j'}(t)D_j^{\dagger}(t')D_l^{\dagger}(t'')D_i(t''')|0\rangle \theta(t'''-t'')\theta(t''-t')\theta(t'-t)\\[0.15cm]
        & = \text{exp}\left[\lambda^2 \delta_{j'j}e^{(i\omega_{\text{vib}}-\Gamma)(t'-t)}\right] 
        \text{exp}\left[\lambda^2 \delta_{il}e^{(i\omega_{\text{vib}}-\Gamma)(t'''-t'')}\right]
        \text{exp}\left[-\lambda^2 \delta_{jl}e^{(i\omega_{\text{vib}}-\Gamma)(t''-t')}\right]\\[0.15cm]
        & \quad \times \text{exp}\left[\lambda^2 \delta_{j'l}e^{(i\omega_{\text{vib}}-\Gamma)(t''-t)}\right]
        \text{exp}\left[\lambda^2 \delta_{ij}e^{(i\omega_{\text{vib}}-\Gamma)(t'''-t')}\right]
        \text{exp}\left[-\lambda^2 \delta_{ij'}e^{(i\omega_{\text{vib}}-\Gamma)(t'''-t)}\right].
    \end{split}
\end{equation}
For the other 23 segments, the calculations proceed as above by using Eq.(\ref{bnc}). The results will be presented elsewhere. Eq.(\ref{vibcorr})
will be useful for calculating the 3rd-order signals entailed later.

\begin{figure*}[t]
 \captionsetup{justification=raggedright,singlelinecheck=false}
\centering
\includegraphics[scale=0.5]{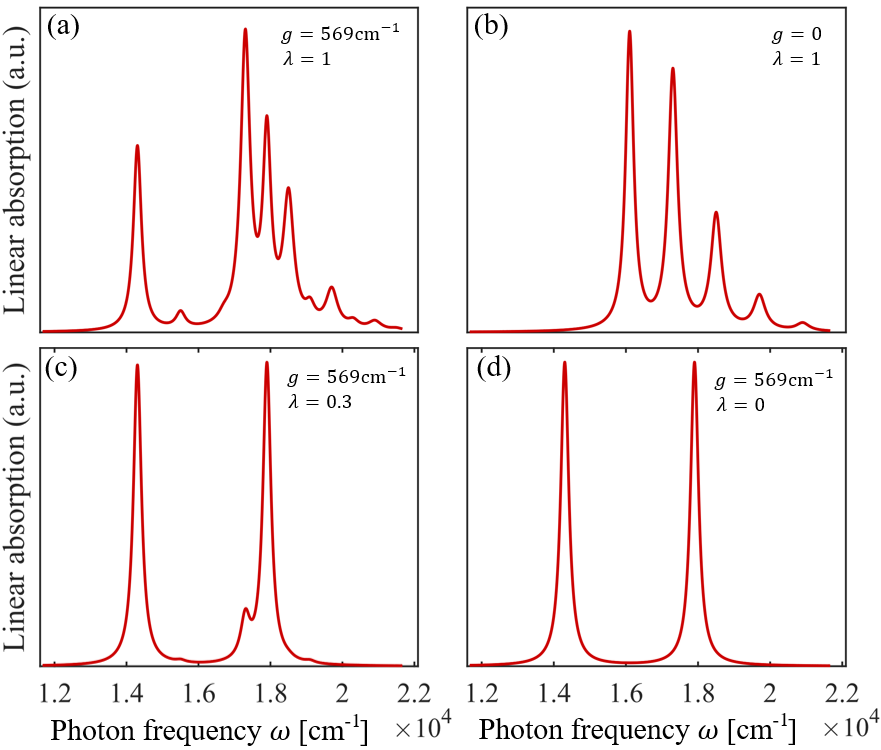}
\caption{Linear absorption spectrum of molecular polaritons with $N=10$ organic molecules in an optical cavity, with different exciton-photon and vibronic couplings. The parameters are taken to be $\omega_n=16113$cm$^{-1}$, $\tilde{\delta}=\delta_c=0$, $\Gamma=20$cm$^{-1}$, $\gamma=1$cm$^{-1}$, $\gamma_c=0.9$cm$^{-1}$, $\omega_{\text{vib}}=1200$cm$^{-1}$, typically from cyanine dyes \cite{Peon_JPCB2013}.}
\label{F2}
\end{figure*}

\section{Absorption and Emission spectra}
Using Eq.(\ref{qlep4}) it is straightforward to find the solution to the 1st order with respect to the external field. The $n_j$ in $\text{W}_\text{x}$ is responsible for the emission whereas the 1 in $\text{W}_\text{x}$ is for absorption. Subject to a probe pulse, Eq.(\ref{qlep4}) yields the 1st-order solution
\begin{equation}
\mathrm{V}_1(t)= i \int_{0}^{t} \Omega\left(\tau-T\right) e^{i v T} G(t-\tau) \mathrm{W}_{\mathrm{x}}(\tau) d\tau.
\label{V1}
\end{equation}
The far-field polarization of cavity polaritons thus reads
\begin{subequations}
    \begin{align}
        & P_A(t) = -i|\mu_{eg}|^2 \sum_{i=1}^N\sum_{l=1}^N\int_{-\infty}^t d\tau E(\tau-T)e^{-iv(\tau-T)} \langle 0|G_{il}(t-\tau) D_i(t) D_l^{\dagger}(\tau)|0\rangle \label{PA} \\[0.15cm]
        & P_E(t) = 2i|\mu_{eg}|^2 \sum_{i=1}^N\sum_{l=1}^N\int_{-\infty}^t d\tau E(\tau-T)e^{-iv(\tau-T)} \langle 0|G_{il}(t-\tau) n_l(\tau) D_l^{\dagger}(\tau)D_i(t)|0\rangle \label{PE}
    \end{align}
\end{subequations}
for absorption and emission components, while the Fourier transform of Eq.(\ref{PE}) gives Eq.(6) in main text. The propagator $G(t)$ is of the same form as the one given in main text. Eq.(\ref{PA}) defines the 1st-order response function $\chi^{(1)}(\omega)$ through the Fourier component of the $P_A(t)$
\begin{equation}
    \begin{split}
        \chi^{(1)}(\omega) = |\mu_{eg}|^2 \left[\sum_{k=1}^{N+1}\frac{\left(\sum_{i=1}^N T_{ik}\right)\left(\sum_{l=1}^N T_{kl}^{-1}\right)e^{-\lambda^2}}{\omega - \omega_k + i\gamma_k} + \sum_{k=1}^{N+1}\sum_{m=1}^{\infty}\frac{\left(\sum_{l=1}^N T_{lk}T_{kl}^{-1}\right)S_m^{\lambda}}{\omega - \omega_k - m\omega_{\text{vib}} + i(\gamma_k+m\Gamma)}\right]
    \end{split}
\label{SA}
\end{equation}
and $S_m^{\lambda}=e^{-\lambda^2}\lambda^{2m}/m!$ is the Franck-Condon factor \cite{Franck_TFS1925,Huang_RPRSA1950}. The 1st term in Eq.(\ref{SA}) indicates the polaritons while the peak vanishes as $k=\text{Dark}$. The 2nd term indicates the emitter dark states (EDSs) decoupled from the cavity photons, responsible for the breakdown of the cooperativity in the cavity polaritons and dark states. To see the absorption closely, we can assume $\gamma=\gamma_c,\ \tilde{\delta}=\delta_c=0$ in the M matrix, which results in $T^{-1}=T^{\dagger}$. We can solve for the $T$ matrix, having $\sum_{l=1}^N T_{lk}=0,\ |T_{lk}|^2=1/N$ for $k=\text{Dark}$ and $\sum_{l=1}^N T_{lk}=\sqrt{N/2},\ \sum_{l=1}^N |T_{lk}|^2=1/2$ for $k=$ LP and UP. For a few certain modes, i.e., $\omega=\omega_{\text{LP}/\text{UP}},\ \omega_{\text{LP}/\text{UP}}+m\omega_{\text{vib}},\ \omega_D+m\omega_{\text{vib}}$, the peak intensities can be found accordingly which yields the ratios
\begin{subequations}
\begin{align}
&\frac{S_{A}\left(\omega_{\mathrm{D}}+m \omega_{\mathrm{vib}}\right)}{S_{A}\left(\omega_{\mathrm{LP} / \mathrm{UP}}\right)}=\frac{\lambda^{2 m}}{m !}\left(1-\frac{1}{N}\right) \frac{2 \gamma_{\mathrm{LP} / \mathrm{UP}}}{\gamma_{\mathrm{D}}+m \Gamma} \approx 2 \frac{\lambda^{2 m}}{m !} \frac{\gamma_{\mathrm{LP} / \mathrm{UP}}}{m \Gamma} \quad \text { as } \quad N \gg 1 \\[0.15cm]
&\frac{S_{A}\left(\omega_{\mathrm{LP} / \mathrm{UP}}+m \omega_{\mathrm{vib}}\right)}{S_{A}\left(\omega_{\mathrm{LP} / \mathrm{UP}}\right)}=\frac{\lambda^{2 m}}{m ! N} \frac{\gamma_{\mathrm{LP} / \mathrm{UP}}}{\gamma_{\mathrm{LP} / \mathrm{UP}}+m \Gamma} \approx 0 \quad \text { as } \quad N \gg 1.
\end{align}
\label{SAratio}
\end{subequations}

Between the two polariton states, we can observe an extra peak at $\omega_D+\omega_{\text{vib}}$ supporting an EDS that results from the vibronic coupling quantified by the Frank-Condon factor. This is evident by Eq.(\ref{SAratio}a), indicating the EDSs decoupled from the cavity photons and the large oscillator strength owing to the density of states $\sim N$ shown in Fig.\ref{F2}(a). The properties of EDSs can be further seen from Fig.\ref{F2}(b) for the case when turning off the cavity, as a benchmark to strong-coupling cases. For weaker vibronic coupling, as shown in Fig.\ref{F2}(c,d), the EDSs are masked by the Rabi splitting that evidences the strong molecular cooperativity. Therefore the EDSs reveal the polariton-induced cooperativity between molecules broken by the vibration fluctuations. As will be shown, the polariton dynamics against the EDSs, for a deeper understanding of molecular cooperativity, can be monitored by the multidimensional projections of the signal.

\section{TWO-DIMENSIONAL COHERENT SIGNAL}
To find the two-dimensional spectroscopic signal, we are seeking the solution in series of the molecule-field interactions, i.e., $\text{V}=\text{V}_0+\text{V}_1+\text{V}_2+\text{V}_3+\cdots$. The 0th-order term $\text{V}_0$ is a result of free propagation in the absence of classical pulses, while the 1st-order term denotes the absorption under the pulse 1, creating the coherence between the ground and electronically excited states. Namely, it reads
\begin{subequations}
  \begin{align}
      & \text{V}_0(t) = G(t)\text{V}(0) + i\int_0^t G(t-t')\text{V}^{\text{in}}(t')dt' \\[0.15cm]
      & \text{V}_1(t) = -i\sum_{s=1}^3\int_0^t dt'\Omega_s(t'-T_s)e^{iv_s T_s}G(t-t')\bar{\text{W}}(t')e^{-i(v_s-v_3)t'}
  \end{align}
\label{V01}
\end{subequations}
and $\bar{\text{W}}(t')=[D_1^{\dagger}(t'),D_2^{\dagger}(t'),\cdots,D_N^{\dagger}(t'),0]^{\text{T}}$, which yield the electronic polarization to the 1st order
\begin{equation}
    \begin{split}
        \tilde{\sigma}_{l;1}^-(t) = -i\sum_{s=1}^3\sum_{j=1}^N \int_0^t dt' \Omega_s(t'-T_s)e^{iv_s T_s}G_{lj}(t-t')D_j^{\dagger}(t')e^{-i(v_s-v_3)t'}.
    \end{split}
\label{sigma1}
\end{equation}

Knowing that the excitations are produced by the first two pulses, one has $n_{l;2}(t)=\sigma_{l;1}^+(t)\sigma_{l;1}^-(t)$, yielding
\begin{equation}
   \begin{split}
       n_{l;2}(t) = \sum_{j=1}^N\sum_{j'=1}^N\int_0^t dt'' & \int_0^t dt' \Omega_1^*(t'-T_1)\Omega_2(t''-T_2) e^{-i(v_1 T_1-v_2 T_2)} e^{-i(v_3-v_1)t'}e^{i(v_3-v_2)t''}\\[0.15cm]
       & \times G_{lj'}^*(t-t')G_{lj}(t-t'')D_{j'}(t')D_j^{\dagger}(t'') + \left[\text{Other\ combinations}\right].
   \end{split}
\label{n2}
\end{equation}
It is obvious that the 2nd-order correction $\text{V}_2(t)$ must involve the noise operators in the form of $\tilde{\sigma}_i^{\pm,\text{in}}(t)$ and $\tilde{\sigma}_i^{+,\text{in}}(t)\tilde{\sigma}_j^{-,\text{in}}(t')\tilde{\sigma}_k^{-,\text{in}}(t'')$ + their Hermitian conjugates. Proceeding to the 3rd-order correction, the noise-operator-dependent terms remain the combinations $\tilde{\sigma}_i^{+,\text{in}}(t)\tilde{\sigma}_j^{-,\text{in}}(t')\tilde{\sigma}_k^{+,\text{in}}(t'')\tilde{\sigma}_l^{-,\text{in}}(t''')$ as well as $\tilde{\sigma}_i^{+,\text{in}}(t)\tilde{\sigma}_j^{+,\text{in}}(t')\tilde{\sigma}_k^{-,\text{in}}(t'')\tilde{\sigma}_l^{-,\text{in}}(t''')$. These will not contribute to the 3rd-order signals eventually, by noting
\begin{equation}
  \begin{split}
    \langle \tilde{\sigma}_i^{+,\text{in}}(t)\tilde{\sigma}_j^{-,\text{in}}(t')\tilde{\sigma}_k^{+,\text{in}}(t'')\tilde{\sigma}_l^{-,\text{in}}(t''') \rangle = 0,\quad \langle \tilde{\sigma}_i^{+,\text{in}}(t)\tilde{\sigma}_j^{+,\text{in}}(t')\tilde{\sigma}_k^{-,\text{in}}(t'')\tilde{\sigma}_l^{-,\text{in}}(t''') \rangle = 0
  \end{split}
\end{equation}
given $\langle \tilde{\sigma}_i^{+,\text{in}}(t)\tilde{\sigma}_j^{-,\text{in}}(t')\rangle=0$. This is reasonable because creating the excitons via environmental fluctuations is of extremely low chance, at room temperature. Keeping the significant components for the signals, we have $\text{V}_2(t)=0$ therein, by noting the fact $\text{V}(0)=0$ prior to the pulse action. We therefore proceed to the 3rd-order correction via Eq.(\ref{qlep4}) and (\ref{n2}), obtaining
\begin{equation}
    \begin{split}
        \text{V}_3(t) = 2i\int_0^t d\tau\  \Omega_3(\tau-T_3) e^{iv_3 T_3}G(t-\tau)\text{W}_{\text{y}}(\tau) + \left[\text{Other\ combinations}\right]
    \end{split}
\label{V3}
\end{equation}
where $\text{W}_{\text{y}}(\tau)=[n_{1;2}(\tau)D_1^{\dagger}(\tau),n_{2;2}(\tau)D_2^{\dagger}(\tau),\cdots,n_{N;2}(\tau)D_N^{\dagger}(\tau),0]^{\text{T}}$. As the three pulses are time-ordered, the transition pathways are selective. This enables us to simplify the 3rd-order polarization of molecules, resulting from the term cancellation. Along the direction $\textbf{k}_{\text{I}}=-\textbf{k}_1+\textbf{k}_2+\textbf{k}_3$, the far-field dipolar radiation reads from Eq.(\ref{V3})
\begin{equation}
    \begin{split}
        \sigma_{i;3}^-(t) = 2i\sum_{l=1}^N\int_0^t d\tau\ \Omega_3(\tau-T_3)e^{-iv_3(t-T_3)}G_{il}(t-\tau)n_{l;2}(\tau)D_l^{\dagger}(\tau)D_i(t).
    \end{split}
\label{sigma3}
\end{equation}
So far, we have derived in a math rigor the 3rd-order polarization of molecules, keeping all the essential terms. Thus Eq.(\ref{sigma3}) must contain the transient absorption component, provided that the eigenvalues of the matrix $\hat{\text{M}}$ normally involve the frequencies corresponding to the higher transition ladders, i.e., $e\rightarrow f$. This is an important aspect for a self-consistent formalism of the spectroscopic signals, as the ultrashort pulses having broad bandwidth would induce considerably the transitions to the doubly-excited manifold. For instance, a large number of identical molecules that may enable the approximation $g\sigma_l^z\approx g$ when including up to the doubly-excited manifold (applicable for few-photon cavities), the cascading transitions $g\rightarrow e,\ e\rightarrow f$ will have the same frequencies. This can be seen from the eigenvalues of the $\hat{\text{M}}$, given no direct many-particle couplings between molecules.

Inserting Eq.(\ref{n2}) into Eq.(\ref{sigma3}) and carrying out the Fourier transform we obtain the macroscopic far-field polarization $P(t)=\mu_{eg}^*\sum_{i=1}^N\langle \sigma_{i;3}^-(t)\rangle$ which yields
\begin{equation}
    \begin{split}
        P(\omega) & = 2i\mu_{eg}^* \sum_{i,l=1}^N\sum_{j,j'=1}^N\sum_{m=1}^{N+1} \int_{-\infty}^{\infty}dt \int_0^t d\tau \int_0^{\tau}dt'' \int_0^{t''}dt'\  e^{i\omega T_3} e^{i(\omega-v_3)(t-T_3)} e^{iv_3(t''-t')} e^{-iv_2(t''-T_2)}  \\[0.15cm]
        & \qquad\qquad\qquad\qquad\quad \times e^{iv_1(t'-T_1)} \Omega_3(\tau-T_3) \Omega_2(t''-T_2) \Omega_1^*(t'-T_1) \langle 0| G_{il}(t-\tau) G_{lm}^*(\tau-t'')\\[0.15cm]
        & \qquad\qquad\qquad\qquad\quad \times G_{lj}(\tau-t'') G_{mj'}^*(t''-t') D_{j'}(t')D_j^{\dagger}(t'')D_l^{\dagger}(\tau)D_i(t)|0\rangle\\[0.2cm]
        & = 2i\mu_{eg}^*\Omega_3\Omega_2\Omega_1^*\sum_{i,l=1}^N\sum_{j,j'=1}^N\sum_{m=1}^{N+1} \int_{T_3}^{\infty}dt\ e^{i\omega T_3} e^{i(\omega-v_3)(t-T_3)}e^{iv_3(T_2-T_1)} \langle 0|G_{il}(t-T_3)G_{lm}^*(T_3-T_2)\\[0.15cm]
        & \qquad\qquad\qquad\qquad\quad \times G_{lj}(T_3-T_2) G_{mj'}^*(T_2-T_1) D_{j'}(T_1)D_j^{\dagger}(T_2)D_l^{\dagger}(T_3)D_i(t)|0\rangle
    \end{split}
\label{Pw}
\end{equation}
where the last step invokes the {\it impulsive approximation}, i.e., $\Omega_i(t-T_i)\simeq\Omega_i\delta(t-T_i)$ \cite{Mukamel_book1999,Yan_JCP1990}. Using Eq.(\ref{bnc}) and (\ref{vibcorr}), the vibrational correlation function can be evaluated
\begin{equation}
    \begin{split}
        & \langle 0|D_{j'}(T_1)D_j^{\dagger}(T_2)D_l^{\dagger}(T_3)D_i(t)|0\rangle \\[0.15cm]
        & = e^{-2\lambda^2}\text{exp}\left[\lambda^2\delta_{j'j}e^{(i\omega_{\text{vib}}-\Gamma)(T_2-T_1)}\right] \text{exp}\left[\lambda^2\delta_{il}e^{(i\omega_{\text{vib}}-\Gamma)(t-T_3)}\right] \text{exp}\left[-\lambda^2\delta_{jl}e^{(i\omega_{\text{vib}}-\Gamma)(T_3-T_2)}\right]\\[0.15cm]
        & \quad \times \text{exp}\left[\lambda^2\delta_{j'l}e^{(i\omega_{\text{vib}}-\Gamma)(T_3-T_1)}\right] \text{exp}\left[\lambda^2\delta_{ij}e^{(i\omega_{\text{vib}}-\Gamma)(t-T_2)}\right] \text{exp}\left[-\lambda^2\delta_{ij'}e^{(i\omega_{\text{vib}}-\Gamma)(t-T_1)}\right]\\[0.2cm]
        & = e^{4\lambda^2}\sum_{m_1=0}^{\infty}\sum_{m_2=0}^{\infty}\sum_{m_3=0}^{\infty}\sum_{m_4=0}^{\infty}\sum_{m_5=0}^{\infty}\sum_{m_6=0}^{\infty}S_{m_1}^{\lambda}S_{m_2}^{\lambda}S_{m_3}^{\lambda}S_{m_4}^{\lambda}S_{m_5}^{\lambda}S_{m_6}^{\lambda}(-1)^{m_3+m_6}\delta_{j'j}^{m_1}\delta_{il}^{m_2}\delta_{jl}^{m_3}\delta_{j'l}^{m_4}\delta_{ij}^{m_5}\delta_{ij'}^{m_6}\\[0.15cm]
        & \qquad\qquad \times e^{(m_2+m_5+m_6)(i\omega_{\text{vib}}-\Gamma)(t-T_3)} e^{(m_3+m_4+m_5+m_6)(i\omega_{\text{vib}}-\Gamma)(T_3-T_2)} e^{(m_1+m_4+m_6)(i\omega_{\text{vib}}-\Gamma)(T_2-T_1)}
    \end{split}
\label{VC}
\end{equation}
with $T_1<T_2<T_3<t$. Inserting Eq.(\ref{VC}) into Eq.(\ref{Pw}), the polarization can be obtained. The heterodyne-detected signal is $S(\omega,T,\omega_e)=\int_0^{\infty}E_{\text{LO}}^*(\omega)P(\omega) e^{i\omega_e\tau}d\tau$, where $E_{\text{LO}}(\omega)$ is the Fourier component of the reference beam. Thus
%The reference pulse serves as a local oscillator interfering with the emission. This leads to the heterodyne signal such that $S(\omega,T,\omega_e)=\int_0^{\infty}E_{\text{LO}}^*(\omega)P(\omega) e^{i\omega_e\tau}d\tau$ with carrying out the Fourier transform against the delay $\tau=T_2-T_1$, where $E_{\text{LO}}(\omega)$ is the Fourier component of the local oscillator field. The result reads
\begin{equation}
   \begin{split}
       S(\omega,T,\omega_e) = & -i|\mu_{eg}|^4 E_{\text{LO}}^* E_3 E_2 E_1^* \sum_{i,l=1}^N\sum_{j,j'=1}^N\sum_{p=1}^{N+1}\sum_{k_1,k_2=1}^{N+1}\sum_{\{m\}=0}^{\infty} e^{4\lambda^2} S_{m_1}^{\lambda}S_{m_2}^{\lambda} S_{m_3}^{\lambda}S_{m_4}^{\lambda}S_{m_5}^{\lambda}S_{m_6}^{\lambda} (-1)^{m_3+m_6}\\[0.15cm]
       & \ \times \frac{\delta_{j'j}^{m_1}\delta_{il}^{m_2}\delta_{jl}^{m_3}\delta_{j'l}^{m_4}\delta_{ij}^{m_5}\delta_{ij'}^{m_6} T_{i,k_1}T_{k_1,l}^{-1}}{\omega-\omega_{k_1}+(m_2+m_5+m_6)\omega_{\text{vib}}+i[\gamma_{k_1}+(m_2+m_5+m_6)\Gamma]} G_{lp}^*(T)G_{lj}(T)\\[0.15cm]
       & \ \times e^{(m_3+m_4+m_5+m_6)(i\omega_{\text{vib}}-\Gamma)T}\frac{T_{p,k_2}^* T_{k_2,j'}^{-1,*}}{\omega_e+\omega_{k_2}+(m_1+m_4+m_6)\omega_{\text{vib}}+i[\gamma_{k_2}+(m_1+m_4+m_6)\Gamma]}.
   \end{split}
\label{S2D}
\end{equation}
The signal measured in detectors is $S_{\text{2D}}(\omega,T,\omega_e)=\text{Im}[S(\omega,T,\omega_e)]$.

\section{ULTRAFAST PUMP-PROBE SIGNAL}
To monitor the polariton dynamics of molecular ensembles, a pump-probe technique composing two broadband pulses with a time delay may be an alternative signal. This acquires the far-field polarization involving the 3rd-order expansion in the field-molecule interaction, as what has been done for the multidimensional signal. The procedures emulate the one for the multidimensional signal entailed before, so that we will present the most significant terms to avoid redundancy, with the rigorous math given in previous section. Invoking the {\it impulsive} approximation in regard of the ultrashort pulses, i.e., $\Omega_i(t-T_i)\approx \Omega_i \delta(t-T_i)$, the vibration correlation function is found to be
\begin{equation}
    \begin{split}
        \langle 0|D_{j'}(T_1)D_j^{\dagger}(T_1)D_l^{\dagger}(T_3)D_i(t)|0\rangle & = e^{-2\lambda^2} \text{exp}\left[\lambda^2\delta_{il}e^{(i\omega_{\text{vib}}-\Gamma)(t-T_3)}\right] \text{exp}\left[\lambda^2(\delta_{j'l} - \delta_{jl})e^{(i\omega_{\text{vib}}-\Gamma)(T_3-T_1)}\right]\\[0.15cm]
        & \qquad \times\text{exp}\left[\lambda^2(\delta_{ij} - \delta_{ij'})e^{(i\omega_{\text{vib}}-\Gamma)(t-T_1)}\right]\\[0.2cm]
        & = e^{\lambda^2} \sum_{m_1}^{\infty} \sum_{m_2}^{\infty} \sum_{m_3}^{\infty} S_{m_1}^{\lambda} S_{m_2}^{\lambda} S_{m_3}^{\lambda} \delta_{il}^{m_1} \left(\delta_{j'l} - \delta_{jl}\right)^{m_2} \left(\delta_{ij} - \delta_{ij'}\right)^{m_3}\\[0.15cm]
        & \qquad\qquad\qquad \times e^{(m_1+m_3)(i\omega_{\text{vib}}-\Gamma)(t-T_3)} e^{(m_2+m_3)(i\omega_{\text{vib}}-\Gamma)(T_3-T_1)}
    \end{split}
\label{VCpp}
\end{equation}
where $T_1$ and $T_3$ denote the arrival time of pump and probe pulses, respectively. The time delay follows $T=T_3-T_1$. The macroscopic polarization can be found correspondingly, through substituting Eq.(\ref{VCpp}) into (\ref{Pw}). The signal measured in pump-probe experiments is the transmission of the probe field with a time grating, i.e., $S_{\text{pp}}(\omega,T)=\text{Im}[E_3^*(\omega)P(\omega)]$. Accounting for the non-rephasing component with similar manipulations, one reaches the full pump-probe signal
\begin{equation}
    \begin{split}
        S_{\text{pp}}(\omega,T) = -4 |\mu_{eg}|^4 & |E_3|^2  |E_1|^2 \sum_{i,l=1}^N\sum_{j,j'=1}^N\sum_{k=1}^{N+1}\sum_{m_1=0}^{\infty}\sum_{m_2=0}^{\infty}\sum_{m_3=0}^{\infty} e^{\lambda^2} S_{m_1}^{\lambda} S_{m_2}^{\lambda} S_{m_3}^{\lambda}\delta_{il}^{m_1} \left(\delta_{j'l} - \delta_{jl}\right)^{m_2}\\[0.2cm]
        & \quad \times \left(\delta_{ij} - \delta_{ij'}\right)^{m_3} \text{Im}\left[\frac{T_{ik}T_{kl}^{-1} G_{lj'}^*(T)G_{lj}(T) e^{(m_2+m_3)(i\omega_{\text{vib}}-\Gamma)T}}{\omega-\omega_k+(m_1+m_3)\omega_{\text{vib}}+i(\gamma_k+(m_1+m_3)\Gamma)}\right].
    \end{split}
\label{SPP}
\end{equation}

The ultrafast pump-probe signal may reveal the collective dynamics of cavity polaritons via simulating Eq.(\ref{SPP}). This is straightforward but is not quite necessary, as the pump-probe signal has simpler form from which some generic features can be observed for polariton dynamics. We will work for the sake of generality, to manifest the the time-resolved signal generically. 
%The ultrafast pump-probe signal for molecular polaritons is illustrated in Fig.4. 
The broadband nature of the ultrashort pulses causes no selective mode absorption in molecular polaritons, whereas the time grating makes the emission spectrally resolved. At zero delay, Eq.(\ref{SPP}) resolves the LP and UP modes separated by $2g\sqrt{N}$, and also the EDSs at $\omega=\omega_D-\omega_{\text{vib}}, \omega_D-2\omega_{\text{vib}}, ...$. The peak intensity distribution over the modes follows the absorption spectrum in Eq.(\ref{SA}), apart from the Stokes shifts of $(negative\ integer) \times\omega_{\text{vib}}$ instead of the anti-Stokes shifts. 
%The Stokes shifts of integer$\times\omega_{\text{vib}}$ are evident by comparing to the absorption spectrum in Fig.\ref{F2}. 
This takes the rational from recalling the absorption and fluorescence spectrums in molecular spectroscopy \cite{Mukamel_book1999,Yan_PRA1990}. By varying the time delay such that $T>0$, the fast relaxation towards the EDSs can be monitored from the dramatic change of the peak intensities at $\omega=\omega_{\text{D}}-m\omega_{\text{vib}}$, i.e.,
\begin{subequations}
    \begin{align}
        & S_{\text{pp}}(\omega_{\text{UP}},T) \propto \frac{e^{-\lambda^2}}{2\gamma_{\text{UP}}} \sum_{l=1}^N\sum_{j,j'=1}^N\sum_{m=0}^{\infty} S_m^{\lambda}\left(\delta_{j'l}-\delta_{jl}\right)^m \text{Re}\left[G_{lj'}^*(T)G_{lj}(T) e^{m(i\omega_{\text{vib}}-\Gamma)T}\right] \label{SppSa}\\[0.2cm]
        & S_{\text{pp}}\left(\omega_{\text{D}}-(m_1+m_3)\omega_{\text{vib}},T\right) \propto \frac{e^{\lambda^2}}{N} \sum_{s,l=1}^N\sum_{j,j'=1}^N\sum_{m_2=0}^{\infty} \frac{S_{m_1}^{\lambda}S_{m_2}^{\lambda}S_{m_3}^{\lambda} \delta_{sl}^{m_1} (\delta_{j'l}-\delta_{jl})^{m_2} (\delta_{sj}-\delta_{sj'})^{m_3}}{\gamma_k + (m_1+m_3)\Gamma}\nonumber\\[0.15cm]
        & \qquad\qquad\qquad\qquad\qquad\qquad\qquad\qquad\quad \times \text{Re}\left[e^{i(\phi_{sk}+\phi_{kl})}G_{lj'}^*(T) G_{lj}(T) e^{(m_2+m_3)(i\omega_{\text{vib}}-\Gamma)T}\right]. \label{SppSb}
    \end{align}
\end{subequations}
Eq.(27) evidences the coherent energy transfer and exchange between polariton states and EDSs, from the phase factor $e^{i(\phi_{sk}+\phi_{kl})}$ in Eq.(\ref{SppSb}) associated with different damping rates that are responsible for the incoherent channels of relaxation of EDSs. $\phi_{sk}=e^{-2\pi i s(k-1)/N}$ for molecules placed in a cavity with a linear fashion along the cavity axis. One would expect in further the $\pi$ phase difference between the time-evolving dynamics of UP/LP and EDSs, for properly-chosen parameters. Because of the density of states $\sim N$ for EDSs, higher than the one for polaritons, the EDSs show the localized nature. The polariton states featuring the extended waves are therefore trading off with the EDS-induced localization, as indicated from Eq.(\ref{SppSa}) and Eq.(\ref{SppSb}) showing multiple channels and timescales for the relaxation processes. The pump-probe signal, however, cannot provide more detailed information in this regard. 
%This is further supported by the $\pi$ phase difference between the oscillations of the peaks at UP/LP and EDSs, through zooming in on the time-resolved dynamics of the signal. 
Advanced information about the polariton dynamics against the EDSs, for a deeper understanding of molecular cooperativity and relaxations of polaritons and dark states, can be learned from extending the pump-probe scheme to multidimensional projections of the signal.